\documentclass[10pt,leqno]{amsart}

\usepackage[utf8]{inputenc}
\usepackage[T1]{fontenc}
\usepackage{amsmath,amssymb,amsthm,amsfonts}
\usepackage{graphicx}
\usepackage{xcolor}
\usepackage{booktabs}
\usepackage{array}
\usepackage{makecell}
\usepackage{booktabs}
\usepackage{multirow}
\usepackage{rotating}
\usepackage{ragged2e} 
\usepackage{makecell}
\usepackage{float}
\usepackage{textcomp}
\usepackage{url}
\usepackage{verbatim}
\usepackage{multicol}
\usepackage{microtype}
\usepackage{tikz}
\usetikzlibrary{patterns}
\usepackage{pgfplots}
\usepackage{pgfplotstable}
\pgfplotsset{compat=1.7}
\usepackage{subfig} 
\usepackage{algorithm}
\usepackage{algorithmic}
\usepackage{csquotes}
\usepackage[numbers,sort&compress]{natbib}          
\usepackage{epstopdf}       
\usepackage[strings]{underscore} 
\usepackage[hidelinks]{hyperref} 

\usepackage[utf8]{inputenc}
\numberwithin{equation}{section}
\DeclareUnicodeCharacter{0080}{ }
\UseRawInputEncoding
\usepackage[utf8]{inputenc}
\usepackage[T1]{fontenc}
\usepackage{microtype}  
\usepackage[english]{babel}

\begin{document}
\title{Transforming Multi-Omics Integration with GANs: Applications in Alzheimer’s and Cancer} 

\author[Reza]{Md.~Selim Reza}
\address{Tulane Center for Biomedical Informatics and Genomics, Deming Department of Medicine, Tulane University, New Orleans, LA 70112, USA}
\email{mreza@tulane.edu}

\author[unicamp]{Sabrin Afroz}

\author[Rahman]{Md.~Mostafizer Rahman}
\address{Department of Computer Science, Tulane University, New Orleans, LA, USA}
\email{mrahman9@tulane.edu}
\email{mostafiz26@gmail.com}

\author[Alam]{Md.~Ashad Alam}
\address{Ochsner Center for Outcomes Research, Ochsner Research, New Orleans, LA 70121, USA}
\email{mdashad.alam@ochsner.org}

\begin{abstract}
Multi-omics data integration is crucial for understanding complex diseases, yet limited sample sizes, noise, and heterogeneity often reduce predictive power. To address these challenges, we introduce \textit{Omics\_GAN}, a Generative Adversarial Network (GAN)-based framework designed to generate high-quality synthetic multi-omics profiles while preserving biological relationships. We evaluated \textit{Omics\_GAN} on three omics types (mRNA, miRNA, and DNA methylation) using the ROSMAP cohort for Alzheimer’s disease (AD) and TCGA datasets for colon and liver cancer. A support vector machine (SVM) classifier with repeated 5-fold cross-validation demonstrated that synthetic datasets consistently improved prediction accuracy compared to original omics profiles. The AUC of SVM for mRNA improved from 0.72 $\pm$ 0.05 to 0.74 $\pm$ 0.03 in AD, and from 0.68 $\pm$ 0.04 to 0.72 $\pm$ 0.05 in liver cancer. Synthetic miRNA enhanced classification in colon cancer from 0.59 $\pm$ 0.05 to 0.69 $\pm$ 0.03, while synthetic methylation data improved performance in liver cancer from 0.64 $\pm$ 0.06 to 0.71 $\pm$ 0.05. Boxplot analyses confirmed that synthetic data preserved statistical distributions while reducing noise and outliers. Feature selection identified significant genes overlapping with original datasets and revealed additional candidates validated by GO and KEGG enrichment analyses. Finally, molecular docking highlighted potential drug repurposing candidates, including Nilotinib for AD, Atovaquone for liver cancer, and Tecovirimat for colon cancer. \textit{Omics\_GAN} enhances disease prediction, preserves biological fidelity, and accelerates biomarker and drug discovery, offering a scalable strategy for precision medicine applications. 
\end{abstract} 

\maketitle

\section{Introduction}
\label{sec:related}
Multi-omics analysis approaches are used to improve the classification of diseases that are important for treatment and to identify biomarkers for chronic diseases, including Alzheimer and cancer. It also helps us understand how different types of biological data are connected to these diseases (Chen, 2023). While multi-omics analysis is widely used in human genetics, collecting large datasets and minimizing noisy data present significant challenges to the development of effective learning models (Rio A. Kavulru, 2018). To bridge the gap, our study proposes an effective deep learning approach utilizing Generative Adversarial Networks (GANs).

Early detection and accurate prediction of chronic diseases  related pathology are crucial for developing effective interventions and treatment strategies (Yao 2023;  Thushara 2022,  Greenberg 2020). Applying deep learning algorithms  classify long-term illnesses outcomes is promising but challenging (KHAGI 2020, OH 2019). Developing deep learning algorithms that can process complex data, integrating multiple biological data, and provide accurate predictions is an ongoing research area \cite{ Ashad-14T, Ashad-11}. Despite this, the presence of noise, including outliers and adversarial attacks, poses a significant hurdle for the development of robust deep learning methods ; Several challenges are notable: (i) small sample sizes and high data complexity hinder model training, (ii) multi-omics noise can mislead deep learning models, and (iii) interpretability is often lacking, hindering insights  (Iqbal T and Ali H in 2018, Wolterink JM in 2017, and Vey BL and Gichoya JW in 2019.).While omics GAN models have been proposed extensively, they are often limited to handling only two omics data types \cite{Ashad-15,Alam-18C}. This restriction poses challenges in fully leveraging the potential of multi-omics data integration, which is crucial for understanding complex long-term illnesses. Expanding GAN models to incorporate multiple omics data types could significantly enhance their ability to uncover intricate patterns and relationships in diverse datasets \cite{Ashad-08, Ashad-13, Alam-16b, Richfield-17}.  To address these existing gaps in the field, our overarching hypothesis revolves around the effective development of adversarial deep learning approaches.

In this work, our overarching goal is to develop an effective adversarial deep learning approach. The proposed method characterizes multi-omics data and their associated networks, enabling robust prediction of disease outcomes using GANs. Essentially, GANs construct synthetic datasets by aligning the distribution of random noise with that of the original data. The process does not involve retaining information from the noise itself but rather matching its distribution appropriately with that of the original data. Our proposed pipeline integrates three distinct Wasserstein GANs (WGANs) to transform three omics datasets into a novel representation. Each WGAN comprises three layers of Fully Connected Neural Networks (CNNs), which generate a dataset based on a single omics dataset and the normalized adjacency matrix.

The proposed method combines three separate omics datasets—mRNA data, methylation data, and miRNA data—along with their corresponding interaction networks to generate respective omics datasets. This model produces new datasets that resemble real datasets but possess stronger molecular signatures, enabling a better understanding of the biological mechanisms leading to disease states. It also improves disease outcome prediction and facilitates interaction network analysis. Using this methodology, we achieve robust classification accuracy compared to what is possible with the original omics datasets analyzed in isolation. This paper highlights the following key aspects:
\begin{itemize}
\item generating interrelated multi-omics data to achieve better prediction outcomes using deep learning approaches, even when working with small datasets.
\item demonstrating how synthetic data improves the identification of biologically relevant features, particularly for underrepresented conditions.
\item integrating multi-omics data to uncover complex biological networks and enhance understanding of disease mechanisms.
\item exploring genomic biomarker-guided candidate drug agents for the treatment of chronic diseases.
\end{itemize}
The structure of this paper is as follows: Section 2 discusses related works. We introduce the proposed Multi-OmicsGAN model, designed to enhance output quality while effectively utilizing small datasets for chronic diseasesin Section 3. Section 4 focuses on the analysis of experimental results. Finally, Section 5 summarizes the research findings and outlines potential directions for future work.

\section{Related work}
\label{Sec:rew}

In contemporary research, several advanced multi-omics data integration frameworks have been developed, broadly categorized into traditional methods and GAN-based approaches. Traditional methods primarily focus on analyzing multi-omics data or extracting network signatures for distinct phenotypic groups. Conversely, GAN-based methods leverage synthetic data generation to enhance predictive accuracy \cite{ Chakraborty-24, Ballard-24}.

\subsection{Non-generative approaches  }
Multi-Omics Factor Analysis (MOFA is designed to analyze  multi-omics data, identifying coordinated transcriptional and epigenetic changes during cell differentiation. It also captures biological and technical sources of variability. However, MOFA’s linear model structure limits its ability to detect non-linear relationships between features across assays. As a supervised learning framework, iOmicsPASS integrates multi-omics data within the context of biological networks. It extracts network signatures predictive of phenotypic groups, maintaining low prediction error rates. However, this method does not facilitate phenotypic group prediction directly and excludes molecules not represented in user-provided network data, potentially leading to a loss of valuable biological insights \cite{baryshnikova2022iomics, Alam-16a, Alam-16b, Ashad-13, Alam-19}.  Neighborhood-based Multi-Omics Clustering (NEMO) operates by analyzing the local neighborhood of each sample to capture similarity patterns across different omics. Despite its utility, NEMO requires each pair of samples to share at least one common omic for partial data and does not provide direct insights into feature importance nemo. PINSPlus is used for tumor subtype discovery by integrating multiple omics datasets in a single analysis. It identifies both known subtypes and novel subgroups with significant survival differences. The model has demonstrated high performance in validating subtypes using metrics like the Rand Index (RI) and Adjusted Rand Index (ARI) across various mRNA datasets \cite{Ashad-10, Ashad-14T}.

A deep learning approach employing a late integration strategy, MOLI was introduced by Sharifi-Noghabi et al. \cite{huang2019salmon}. The method utilizes modality-specific feedforward neural networks (FNNs) to independently extract features from each omic modality. These features are subsequently combined into a unified multi-omic representation, which serves as input for a classification sub-network aimed at predicting drug responses. Although this approach is straightforward and accounts for the distinct characteristics of each modality, it does not explicitly address potential interactions between the different modalities, which could be a limitation.

SALMON \cite{huang2019salmon} focuses on biological interpretability by either organizing data in biologically meaningful ways or incorporating prior domain knowledge. It uses mRNA-seq and miRNA-seq data to predict Cox regression survival in breast cancer. This is achieved by first performing gene co-expression analysis to create eigengene modules, which reduce the original feature space into biologically meaningful latent features.
Two alternative methods, MiNet \cite{hao2019gene} and DeepOmix \cite{zhao2021deepomix}, incorporate prior biological knowledge into their architectures. MiNet utilizes a neural network design structured to mirror biological systems. It features a multi-omics input layer, followed by a gene layer that links the multi-omics data to corresponding genes, and a pathway layer that connects these genes to their known biological pathways. Similarly, DeepOmix [23] employs a deep neural network (DNN) framework with an input gene layer designed to integrate multi-omics data at the gene level. This is followed by a functional module layer, which leverages biological insights to establish connections between the input gene layer and the functional modules, representing true biological relationships.

MoGCN, introduced by Li et al. \cite{li2022mogcn}, incorporates patient similarity information and employs an intermediate integration strategy. This method integrates multiple modalities into a single representation before classification by using an autoencoder (AE) with multiple encoders and decoders that share a common layer. Peng et al. \cite{peng2022drugresponse} proposed MOFGCN, a knowledge-guided connectivity method designed to predict drug responses in cell lines. This approach constructs a heterogeneous network that incorporates a cell line similarity network, a drug similarity network, and known drug-cell line associations. 

Althubaiti et al. \cite{althubaiti2021deepmocca} developed DeepMOCCA, a model that combines protein-protein interactions (PPIs) with multi-omics data for cancer survival prediction. The model integrates germline and somatic variants, methylation, gene expression, and copy number variants into a graph structure where nodes represent genes and edges denote functional interactions between them.

\subsection{Generative approaches }

Mitra et al. \cite{mitra2020multiview} proposed multi-view neighborhood embedding (MvNE), a method designed to learn a unified probability distribution of samples across multiple omics modalities. This approach generates low-dimensional embeddings that maintain the relationships between samples in the transformed space. 
Zuo et al. \cite{zuo2021deep} introduced the deep cross-omics cycle attention method (DCCA), which takes a unique approach to jointly analyze single-cell multi-omics data for various downstream applications. DCCA begins by encoding each omics modality with separate variational autoencoders (VAEs). It then employs cyclical attention transfer to capture and model associations between the different modalities.

Finally, Generative Adversarial Networks (GANs) have revolutionized multi-omics data integration by overcoming the limitations of traditional methods. By generating synthetic data that mimics real datasets, GANs capture non-linear correlations between omics features, offering versatile applications:  
\begin{itemize}
    \item  Synthetic Data Generation**: GANs are used to generate gene expression data from bulk RNA-seq datasets, capturing complex patterns and relationships.    
\item Biomarker Identification**: Interaction networks derived from multi-omics datasets with GANs enhance the detection of critical biomarkers.  
\item Overcoming Limited Sample Sizes**: GANs generate synthetic data to augment small datasets, improving the reliability and robustness of analyses.  
\end{itemize}
These capabilities have made GANs a powerful tool in advancing multi-omics research.

\section{Materials and methods}
\label{sec:materials}
\subsection{Datasets and Networks}
To assess the effectiveness of our approach, we applied our model to datasets from The Cancer Genome Atlas (TCGA) for liver and colon cancer, as well as the Religious Orders Study/Memory and Aging Project (ROSMAP) dataset for Alzheimer's Disease (AD). Initially, we implemented the recommended architecture for biomedical classification tasks using the ROSMAP dataset to distinguish between AD patients and normal controls (NC). This analysis incorporated three distinct omics data types: Omics1 (\textit{mRNA expression}), Omics2 (\textit{DNA methylation}), and Omics3 (\textit{miRNA expression}). The integration of these omics layers aimed to provide a comprehensive and complementary perspective on disease mechanisms. The dataset was sourced from the MOGONET study and is publicly available in the MOGONET ROSMAP Dataset repository. Only samples with complete data across all three omics types were included in our analysis, yielding a total of 182 AD patient samples and 169 NC samples, each characterized by 200 features per omics data type.

Furthermore, we evaluated our method using TCGA datasets for colon and liver cancers, available through the TCGA Studied Cancers repository. Gene expression (\textit{mRNA}), microRNA expression (\textit{miRNA}), and DNA methylation (\textit{Methy}) data for these cancer types were retrieved from the TCGA Multi-Omic Benchmark Dataset. For colon cancer, we analyzed 551 samples, including 121 cancer patient samples and 430 NC samples. For liver cancer, a total of 404 samples were examined, consisting of 172 cancer patient samples and 232 NC samples. In both cases, the Wilcoxon rank test was employed to extract the 500 most significant features across the three omics data types.

To model the interactions between omics data types, we incorporated networks derived from \textit{TargetScanHuman} \cite{PMID29925568}. This database identifies effective miRNA--mRNA interactions using the context++ model, thereby enabling the construction of regulatory networks involving miRNAs. miRNAs play a crucial role in gene regulation by facilitating the degradation of target mRNA molecules, thereby reducing protein translation. Additionally, miRNAs can regulate the expression of DNA methyltransferases, which mediate DNA methylation \cite{PMID37239435}. DNA methylation, particularly at CpG islands in promoter regions, is a well-established mechanism for gene silencing and reduced mRNA transcription \cite{PMID22781841}.A modified adjacency matrix was used to represent these interactions, where a value of $-1$ indicated an interaction between omics data types, while a value of $1$ denoted the absence of interaction. For the ROSMAP dataset, the constructed mRNA--miRNA, mRNA--Methy, and miRNA--Methy bipartite networks comprised 40{,}000 interactions each. Similarly, for TCGA datasets, the mRNA--miRNA, mRNA--Methy, and miRNA--Methy bipartite networks encompassed 250{,}000 interactions each.

\subsection{Background and Notations}
We proposed a multi-omics GAN model, designed to extract information from the inter-omics network. This model integrates omics datasets through a GAN, updating the datasets and generating new feature sets upon successful training. Table \ref{notations} presents the basic notations used in the proposed approach.  The updated expressions, $H_x^{(k)}$ for $Omics_1$, $H_m^{(k)}$ for $Omics_2$, and $H_p^{(k)}$ for $Omics_3$, capture the changes resulting from $k$ updates, where $k \in {1, 2, \ldots, K}$. Table \ref{table:key_functions} provides a concise breakdown of the key functions within the Multi-Omics GAN model, which facilitate the creation of synthetic omics data. Each function serves a specific role during model training and execution. Understanding these functions is crucial for comprehending how OmicsGAN operates to produce synthetic omics data.

\begin{table*}
\centering
\caption{Notations for Multi-Omics GAN Model with mRNA, DNA-methylation, and miRNA}
\label{notations}
\renewcommand{\arraystretch}{1.15}
\setlength{\tabcolsep}{6pt}

\resizebox{\textwidth}{!}{%
\begin{tabular}{>{\centering\arraybackslash}p{4.5cm} 
                >{\centering\arraybackslash}p{11cm}}
\toprule
\textbf{Symbol} & \textbf{Definition} \\
\midrule
$\mathbf{\textit{X}} \in \mathbb{R}^{r \times n}$ & $\text{Omics}_1$ (i.e., \textit{mRNA}) data obtained from gene sequence \\
$\mathbf{\textit{Y}} \in \mathbb{R}^{m \times n}$ & $\text{Omics}_2$ (i.e., \textit{DNA methylation}) data obtained from DNA methylation sequence \\
$\mathbf{\textit{Z}} \in \mathbb{R}^{p \times n}$ & $\text{Omics}_3$ (i.e., \textit{miRNA}) data obtained from miRNA sequence \\
$\mathbf{\textit{h}}_x^{(k)} \in \mathbb{R}^{r \times n}$ & Intermediate value of $\text{Omics}_1$ in the $k^{\text{th}}$ update \\
$\mathbf{\textit{h}}_y^{(k)} \in \mathbb{R}^{m \times n}$ & Intermediate value of $\text{Omics}_2$ in the $k^{\text{th}}$ update \\
$\mathbf{\textit{h}}_z^{(k)} \in \mathbb{R}^{p \times n}$ & Intermediate value of $\text{Omics}_3$ in the $k^{\text{th}}$ update \\
$\mathbf{\textit{H}}_x^{(k)} \in \mathbb{R}^{r \times n}$ & $\text{Omics}_1$ (synthetic) in the $k^{\text{th}}$ update \\
$\mathbf{\textit{H}}_y^{(k)} \in \mathbb{R}^{m \times n}$ & $\text{Omics}_2$ (synthetic) in the $k^{\text{th}}$ update \\
$\mathbf{\textit{H}}_z^{(k)} \in \mathbb{R}^{p \times n}$ & $\text{Omics}_3$ (synthetic) in the $k^{\text{th}}$ update \\
$\mathbf{\textit{F}}_x \in \mathbb{R}^{r \times n}$ & Final $\text{Omics}_1$ (synthetic), $F_x = H_x^{(k^*)}$ \\
$\mathbf{\textit{F}}_y \in \mathbb{R}^{m \times n}$ & Final $\text{Omics}_2$ (synthetic), $F_y = H_y^{(k^*)}$ \\
$\mathbf{\textit{F}}_z \in \mathbb{R}^{p \times n}$ & Final $\text{Omics}_3$ (synthetic), $F_z = H_z^{(k^*)}$ \\
$\mathbf{N} = \text{Omics}_a - \text{Omics}_b,\ a \neq b$ & Adjacency matrix of the interaction network between $\text{Omics}_a$ and $\text{Omics}_b$ \\
$\mathbf{\textit{D}}_x \in \mathbb{R}^{r \times r}$ & Diagonal matrix for mRNA: $\mathbf{\textit{D}}_x(i,i)=\sum_j|\mathbf{\textit{N}}(i,j)|$ \\
$\mathbf{\textit{D}}_y \in \mathbb{R}^{m \times m}$ & Diagonal matrix for DNA methylation: $\mathbf{\textit{D}}_y(i,i)=\sum_j|\mathbf{\textit{N}}(i,j)|$ \\
$\mathbf{\textit{D}}_z \in \mathbb{R}^{p \times p}$ & Diagonal matrix for miRNA: $\mathbf{\textit{D}}_z(i,i)=\sum_j|\mathbf{\textit{N}}(i,j)|$ \\
$\widetilde{S}_{ab} \in \mathbb{R}^{r \times m \times p},\ a,b \in \{1,2,3\}$ & Normalized adjacency matrix, $\widetilde{S}_{ab}=\mathbf{\textit{D}}_a^{-1/2}\mathbf{N}_{ab}\mathbf{D}_b^{-1/2}$ \\
\bottomrule
\end{tabular}%
}
\end{table*}

\begin{table*}[ht]
\centering
\caption{Key Functions in the Multi-OmicsGAN Model}
\label{table:key_functions}
\begin{tabular}{c p{10cm}} 
\hline
\textbf{Function} & \textbf{Description} \\
\hline
\(D_{loss}()\) & Compares intermediate values with updated values during training. \\
\(ReLU()\) & Activates nodes using the weight matrix for progression to the next layer. \\
\(C()\) & Trains the connected neural network for generator training. \\
\(GAN()\) & Generates synthetic data. \\
\hline
\end{tabular}
\end{table*}

\begin{figure*}
    \centering
    \includegraphics[width=\textwidth]{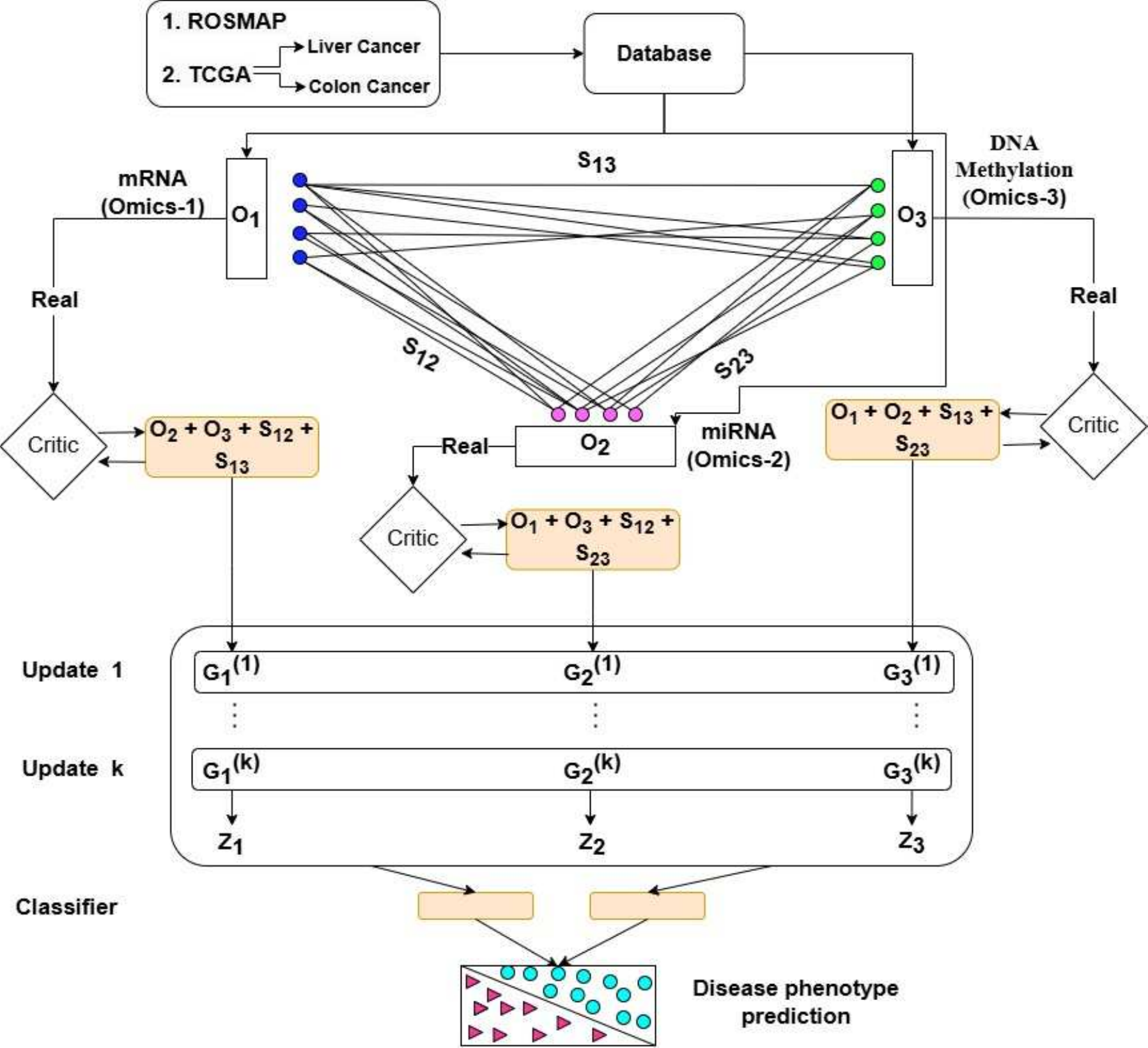} 
    \caption{Predicting the Alzheimer's Disease phenotype utilizing multi-omics datasets through Generative Adversarial Network techniques.}
    \label{fig:3omics}
\end{figure*}

Figure \ref{fig:3omics} illustrates the synthetic data generation process for three omics datasets after the \(k^{th}\) iteration. In each update, the generator first produces an intermediate value for $Omics_a$ using $Omics_b$ (\(a, b \in \{1, 2, 3\} \quad \text{and} \quad a \neq b\)), along with the normalized adjacency matrix $\widetilde{S}$, which represents their interaction network.
\begin{align}
h_x^{(k)} &= GAN(H_y^{(k-1)}, H_z^{(k-1)}, S_{12}, S_{13}) \label{e1} \\
h_y^{(k)} &= GAN(H_z^{(k-1)}, H_x^{(k-1)}, S_{12}, S_{23}) \label{e2} \\
h_z^{(k)} &= GAN(H_x^{(k-1)}, H_y^{(k-1)}, S_{13}, S_{23}) \label{e3}
\end{align}

Equations (\ref{e1}), (\ref{e2}), and (\ref{e3}) describe the generation of synthetic data \( h_x^{(k)}, h_y^{(k)}, \) and \( h_z^{(k)} \) for the \( Omics_1, Omics_2, \) and \( Omics_3 \) datasets at the \( k^{th} \) update. Here, \( H_{x}^{(k-1)}, H_{y}^{(k-1)}, \) and \( H_{z}^{(k-1)} \) represent the updated values of the \( Omics_1, Omics_2, \) and \( Omics_3 \) datasets from the \( (k-1)^{th} \) iteration, while \( S_{ab} \) denotes the interaction network between \( Omics_a \) and \( Omics_b \), processed using the \( GAN() \) function.

\begin{equation}
    \label{e4}
    \text{loss}_{x} = D_{\text{loss}}(h_{x}^{(k)}, H_{x}^{(k-1)})
\end{equation}

\begin{equation}
    \label{e5}
    \text{loss}_{y} = D_{\text{loss}}(h_{y}^{(k)}, H_{y}^{(k-1)})
\end{equation}

\begin{equation}
    \label{e6}
    \text{loss}_{z} = D_{\text{loss}}(h_{z}^{(k)}, H_{z}^{(k-1)})
\end{equation}

The difference between the input and intermediate values is referred to as \(D_{loss}\). During training, this loss [Equations (\ref{e4}, \ref{e5}, \ref{e6})] is minimized by learning the updated datasets \(H_x^{(k)}\), \(H_y^{(k)}\), and \(H_z^{(k)}\) for $Omics_1$, $Omics_2$, and $Omics_3$, respectively. This process ensures that the distributions of \(H_x^{(k)}\), \(H_y^{(k)}\), and \(H_z^{(k)}\) match those of \(H_x^{(k-1)}\), \(H_y^{(k-1)}\), and \(H_z^{(k-1)}\). 
As shown in Figure \ref{fig:3omics}, each update is trained independently. After training, the updated datasets \(H_x^{(k)}\), \(H_y^{(k)}\), and \(H_z^{(k)}\) from update \(k\) become the inputs for the next update, \((k+1)\). The initial inputs to the first layer are \(H_x^{(0)} = X\), \(H_y^{(0)} = Y\), and \(H_z^{(0)} = Z\). After completing the \(k^{th}\) update, the final synthetic datasets used for predicting disease phenotypes are \(F_x = H_x(k^*)\), \(F_y = H_y(k^*)\), and \(F_z = H_z(k^*)\), where \(k^*\) represents the update that yields the best prediction results on a validation dataset.

\subsection{Generative Adversarial Network (GAN)}
In essence, GANs generate synthetic datasets by aligning the distribution of random noise with that of the original data. The goal is not to retain information from the noise itself but to match its distribution to that of the original data \cite{AFROZ2024138}. Our proposed pipeline incorporates three distinct wGANs to transform three omics datasets into a novel representation. Each wGAN consists of three layers of fully connected neural networks (FCNN), which generate a dataset based on a single omics dataset and the normalized adjacency matrix, as defined by the following equations:

\begin{equation}
    \label{e7}
    h_{x}^{(k)}=(ReLU(ReLU(S_{12}S_{13}H_{y}^{(k-1)}H_{z}^{(k-1)}W^{(0)}))W^{(1)})W^{(2)}
\end{equation}

\begin{equation}
    \label{e8}
    h_{y}^{(k)}=(ReLU(ReLU(S_{12}S_{23}H_{z}^{(k-1)}H_{x}^{(k-1)}W^{(0)}))W^{(1)})W^{(2)}.
\end{equation}

\begin{equation}
    \label{e9}
    h_{z}^{(k)}=(ReLU(ReLU(S_{13}S_{23}H_{x}^{(k-1)}H_{y}^{(k-1)}W^{(0)}))W^{(1)})W^{(2)}.
\end{equation}

In Equations (\ref{e7}, \ref{e8}, \ref{e9}), the activation function used is \( ReLU \) (rectified linear unit), and \( W^{l} \) represents the weight matrix in the \( l^{th} \) layer. The intermediate values \( h_x^{(k)} \), \( h_y^{(k)} \), and \( h_z^{(k)} \), along with the input datasets \( H_x^{(k-1)} \), \( H_y^{(k-1)} \), and \( H_z^{(k-1)} \), are obtained by training a fully connected neural network. This network undergoes five rounds of training for each generator, with its training objective defined as:

\begin{equation}
    \label{e10}
    l_{c} = C(h_{x}^{(k)}) - C(H_{x}^{(k-1)}),
\end{equation}

here, \( C \) represents the connected neural network. In this network, real samples (e.g., \( H_x^{k-1} \)) are assigned higher values, while synthetic samples (e.g., \( h_x^{k} \)) are assigned lower values, effectively minimizing Equation \ref{e10}. Synthetic data is generated after five rounds of connected neural network training, with its training objective defined as:

\begin{equation}
    \label{e11}
  l_{G} = -C(h_{x}^{(k)}) + \alpha \|h_{x}^{(k)} - X\|_{2},
\end{equation}
where, \( \alpha \) is a coefficient that determines the weight assigned to the two terms in the equation. Equation \ref{e11} is designed to train the GAN to generate data \( h_x^{k} \) with higher assigned values.

The Generative Adversarial Network (GAN) is a generative modeling technique capable of autonomously identifying patterns in input data without requiring preprocessing or dimensional alterations. Understanding the functionality of the GAN model provides a foundation for the next chapter, which explores experimental analyses involving single-omics, multi-omics, and network-based approaches.

To evaluate our proposed approach, we applied the recommended architecture to biomedical tasks using the ROSMAP dataset, specifically focusing on distinguishing Alzheimer's Disease (AD) patients from normal controls (NC). We leveraged three omics data types: \( Omics_1 \) (mRNA), \( Omics_2 \) (DNA methylation), and \( Omics_3 \) (miRNA). By integrating these data types, we aimed to provide a comprehensive understanding of the diseases. Only samples with complete data for \( Omics_1 \), \( Omics_2 \), and \( Omics_3 \) were considered. The ROSMAP dataset included 182 AD patient samples and 169 NC samples, each characterized by 200 features across the three omics data types.

\subsection{Feature selection }

In this study, we conducted feature selection on omics expression data using the Wilcoxon rank-sum test to identify significant features that differentiate case and control groups. This non-parametric statistical test was selected due to its robustness in handling data with non-normal distributions, making it particularly well-suited for omics data, which often exhibit skewed distributions and the presence of outliers. Features with \textit{p}-values below 0.005 were considered statistically significant. 

The analysis was performed separately on both the original and synthetic datasets, using data from the ROSMAP dataset for AD and the TCGA dataset, which includes liver and colon cancer. After applying the Wilcoxon rank-sum test to both the original and synthetic datasets, we identified the common significant features that appeared consistently across both datasets. These common features were prioritized because they exhibited stable discriminatory power across the original and synthetic datasets, suggesting that they may serve as key driver markers for the diseases of interest. The consistency of these features across original and synthetic datasets increases their reliability as potential biomarkers. Subsequently, these selected common features were considered for further exploration in the context of drug repurposing, with the aim of identifying therapeutic targets that could be leveraged for treating Alzheimer’s disease and various cancer types.

\subsection{Collection of drug agents for exploring candidate drugs}

We conducted a comprehensive literature review focusing on Alzheimer’s disease (AD) to identify potential therapeutic agents guided by host transcriptome data. This process led to the selection of 63 meta-drug candidates that have shown promise in modulating pathways implicated in AD. In parallel, we compiled a library of 2,500 FDA-approved drugs from DrugBank to expand our search for effective treatments targeting liver and colon cancers. These FDA-approved drugs encompass a wide range of pharmacological classes, providing a robust foundation for repurposing opportunities.

Subsequently, we utilized molecular docking analyses to systematically evaluate the binding affinities of the selected compounds with a curated set of target proteins associated with AD, liver cancer, and colon cancer. The 63 meta-drugs for AD and the 2,500 FDA-approved drugs for cancer were screened to identify high-affinity interactions that could suggest potential efficacy in disease modulation. This dual approach, integrating transcriptome-guided selection and comprehensive molecular docking, was aimed at uncovering novel therapeutic candidates with a strong potential for repurposing across these diseases.

\subsection{Drug repurposing by molecular docking study}
In this study, docking analysis was performed between target proteins and approved drugs. The three-dimensional (3D) structures of target proteins were obtained from the Protein Data Bank (PDB) (\url{https://www.rcsb.org/}). In cases where the 3D structures were unavailable in the PDB, the AlphaFold (AF) models were downloaded from UniProt (\url{https://www.uniprot.org/}). The \textit{Discovery Studio Visualizer} (\url{https://www.3ds.com/products/biovia}) was used to visualize the 3D structures of protein interfaces. PDB2PQR and H++ servers were utilized to assign the protonation state of target proteins \cite{PMID17488841, PMID15980491}. All missing hydrogen atoms were also appropriately added. The pKa values for target protein residues were investigated under the following physical conditions: salinity = 0.15, internal dielectric = 10, pH = 7, and external dielectric = 80. On the other hand, the drugs were minimized for energy using Avogadro (\url{https://avogadro.cc/}). The target proteins were solvated with water, and only polar hydrogens were added. The receptor grid boxes (in X, Y, Z dimensions) were prepared in ADT4.2, and the PDBQT files of proteins were generated \cite{morris2009autodock4}. Similarly, the drug agents were prepared with default parameters, and only Gasteiger charges were added. Subsequently, molecular docking between receptors and drug agents was performed to calculate their binding affinities (kcal/mol) using AutoDock Vina \cite{PMID19499576}.  

Flexible ligand docking was performed by applying the Lamarckian Genetic Algorithm with an exhaustiveness value of eight. The contributions of intramolecular hydrogen bonds, hydrophobic, ionic, and Van der Waals interactions between docked protein–ligand complexes were used to determine the free energy ($\Delta G$), specifying affinity scoring of the binding. The docking poses were narrowed down using the force field’s free binding energy computation. After the docked protein–ligand complexes were created, the binding sites were analyzed to construct a 2D representation of the ligand interaction for each complex.

\section{Results}
\label{Sec:res}
\subsection{Omics\_GAN on the ROSMAP and TCGA datasets }
To comprehensively evaluate our generative approach, we conducted experiments on three distinct datasets: the ROSMAP dataset for Alzheimer’s disease (AD) and the TCGA datasets for colon and liver cancers. In this study, we updated the omics profiles (namely, Omics1: gene expression (mRNA), Omics2: miRNA, and Omics3: DNA methylation (Methy)) $K=5$ times to generate synthetic data.  These omics profiles correspond to different biological layers: Omics1 represents mRNA/gene expression data for mRNA–Methy–miRNA interaction networks, Omics2 represents miRNA expression in miRNA–mRNA–Methy interaction networks, and Omics3 represents DNA methylation expression in Methy–miRNA–mRNA interaction networks, as detailed in Table~\ref{table:table3}.  Both the generator and the discriminator in our generative model were implemented as fully connected neural networks with distinct architectures. The generator consisted of two hidden layers with 512 and 768 neurons, respectively, while the discriminator had two hidden layers containing 256 and 128 neurons. For the activation functions, the ReLU activation was applied to the hidden layers of both networks, whereas the output layers employed a linear activation function.  To prevent overfitting, a dropout probability of 0.3 was applied to the hidden layers of the discriminator. Both networks were trained using the RMSprop optimizer with a learning rate of $5 \times 10^{-6}$ for the generator and $5 \times 10^{-5}$ for the discriminator. Additionally, a coefficient $\alpha = 0.01$ was utilized. Hyperparameters were selected based on grid search results, as summarized in Table~\ref{table:table3}.

\begin{figure*}  
    \centering
    \includegraphics[width=\textwidth]{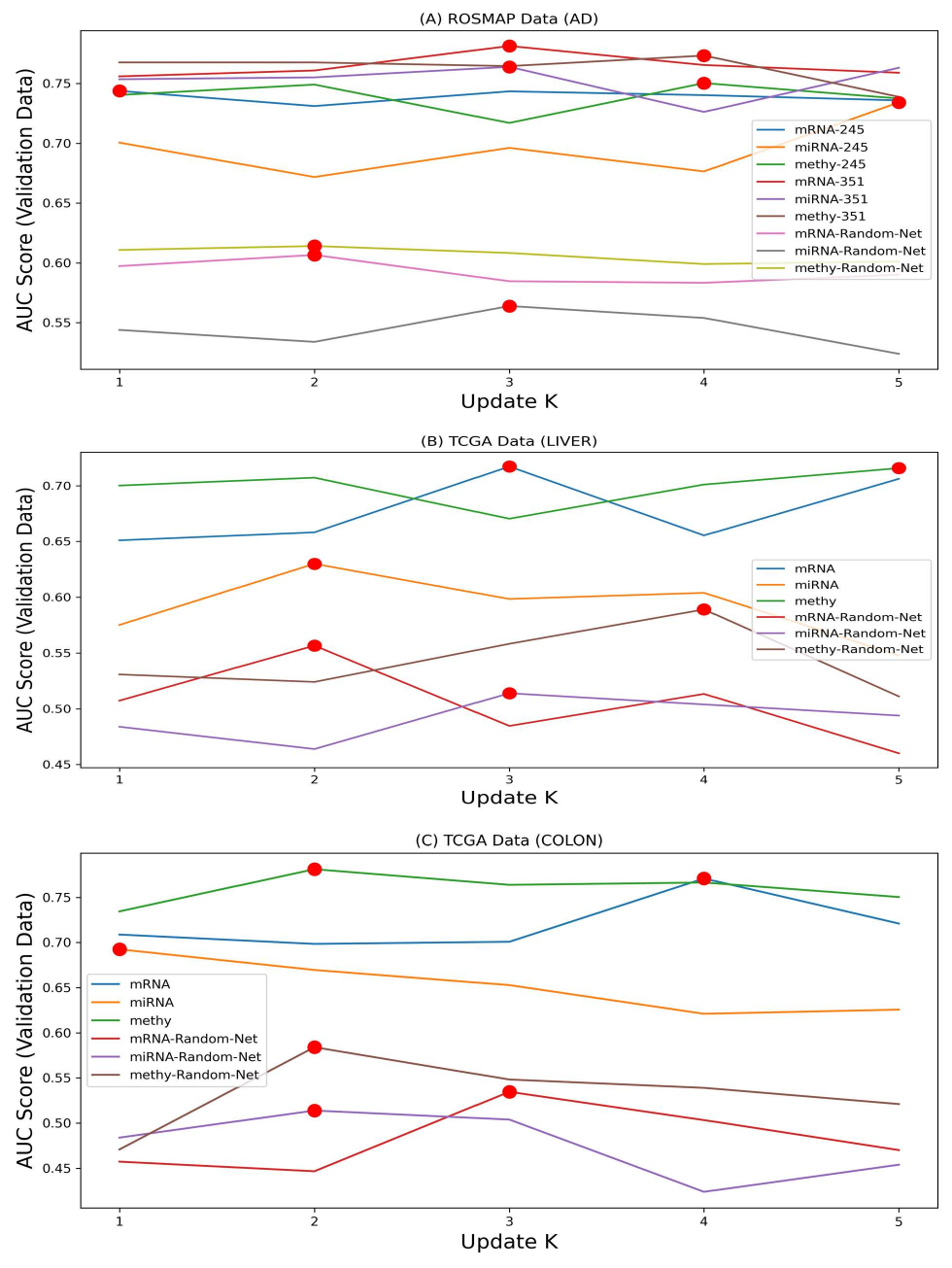} 
    \caption{Prediction results of ROSMAP and TCGA (Liver and Colon cancer) datasets using validation samples. 
        AUC of the prediction results using validation samples of synthetic Omics1, Omics2, and Omics3 with random network for $k = [1, 2, 3, 4, 5]$, 
        where (A) illustrates the ROSMAP dataset with 245 and 351 subset AD patients, 
        (B) illustrates liver cancer, and (C) illustrates colon cancer. 
        Updated $k^*$ with the best validation AUC is selected as the final synthetic data for each omics profile.}  
    \label{fig:Figure2}  
\end{figure*}

The synthetic datasets generated for Omics1, Omics2, and Omics3, updated sequentially for $k = 1,2,\dots,5$, were subsequently fed into a support vector machine (SVM)-based classifier for outcome prediction. As described in the Methods section, the classifier training involved a five-fold cross-validation strategy, wherein each dataset was split into five parts: three folds for training, one fold for validation (for parameter tuning and synthetic data selection), and one fold for testing.  This five-fold split was repeated 50 times per dataset to ensure robustness. To determine the optimal synthetic omics profile for each dataset, we selected the update ($k^{*}$) that yielded the highest area under the curve (AUC) score on the validation set. Specifically, the updated Omics1 profile with the best validation AUC was designated as the final synthetic Omics1 output, and the same approach was applied to Omics2 and Omics3.  Figure~\ref{fig:Figure2} illustrates the process of selecting the final synthetic omics datasets based on validation performance for both the ROSMAP and TCGA datasets.

For the ROSMAP dataset, we analyzed two subsets: one consisting of 245 patients and the other comprising 351 patients. The results indicated that updates $k=1,3$ provided the best validation AUC scores for synthetic Omics1 profiles, while $k=5,3$ achieved the highest validation AUC for synthetic Omics2 profiles for both subsets, respectively. For synthetic Omics3 profiles, $k=4$ offered the best validation performance for both subsets. On the other hand, random networks produced the lowest AUC values for all omics expressions. Therefore, Omics1 updates $1,3$; Omics2 updates $5,3$; and Omics3 update $4$ were used for predicting the test samples, and the corresponding results are reported in this study.  For liver cancer samples from the TCGA dataset (comprising 404 samples), the optimal updates based on validation AUC were as follows: $k=3$ for Omics1, $k=2$ for Omics2, and $k=5$ for Omics3. As observed with the ROSMAP dataset, random networks again produced the lowest AUC scores across all omics profiles. The selected updates—Omics1 update $3$, Omics2 update $2$, and Omics3 update $5$—were used for predicting the test samples.  For the TCGA colon cancer dataset, consisting of 551 samples, the optimal validation AUC scores were achieved with the following updates: $k=2$ for Omics1, $k=1$ for Omics2, and $k=4$ for Omics3. Similar to the previous findings, networks initialized with random parameters resulted in the lowest AUC scores. Consequently, the selected updates—Omics1 update $2$, Omics2 update $1$, and Omics3 update $4$—were used for test sample predictions, and the outcomes are detailed in this study.

\begin{table*}[h!]
    \centering
    \caption{Hyperparameters in Omics GAN used in the study}
    \label{table:table3}
    \renewcommand{\arraystretch}{1}
    \setlength{\tabcolsep}{6pt} 
    \resizebox{\textwidth}{!}{ 
    \begin{tabular}{c c c c}
    \toprule
    \multirow{2}{*}{\textbf{Hyperparameter}} & 
    \multicolumn{3}{c}{\textbf{mRNA–miRNA–methy}} \\
    \cmidrule(lr){2-4}
     & \textbf{ROSMAP} & \textbf{TCGA-Liver} & \textbf{TCGA-Colon} \\
    \midrule
    \textbf{Omics 1 generator learning rate} & $5.00 \times 10^{-6}$ & $5.00 \times 10^{-6}$ & $5.00 \times 10^{-6}$ \\
    \textbf{Omics 1 discriminator learning rate} & $5.00 \times 10^{-5}$ & $5.00 \times 10^{-5}$ & $5.00 \times 10^{-5}$ \\
    \textbf{Omics 1 L2-norm coefficient ($\alpha$)} & $0.01$ & $0.01$ & $0.01$ \\
    \textbf{Omics 2 generator learning rate} & $5.00 \times 10^{-6}$ & $5.00 \times 10^{-6}$ & $5.00 \times 10^{-6}$ \\
    \textbf{Omics 2 discriminator learning rate} & $5.00 \times 10^{-5}$ & $5.00 \times 10^{-5}$ & $5.00 \times 10^{-5}$ \\
    \textbf{Omics 2 L2-norm coefficient ($\alpha$)} & $0.01$ & $0.01$ & $0.01$ \\
    \textbf{Omics 3 generator learning rate} & $5.00 \times 10^{-6}$ & $5.00 \times 10^{-6}$ & $5.00 \times 10^{-6}$ \\
    \textbf{Omics 3 discriminator learning rate} & $5.00 \times 10^{-5}$ & $5.00 \times 10^{-5}$ & $5.00 \times 10^{-5}$ \\
    \textbf{Omics 3 L2-norm coefficient ($\alpha$)} & $0.01$ & $0.01$ & $0.01$ \\
    \bottomrule
    \end{tabular}
    } 
\end{table*}

\subsection{Integration of Omics1, Omics2, and Omics3 }
To explore the potential of integrating multi-omics data for improved disease prediction and biomarker discovery, we generated synthetic Omics1, Omics2, and Omics3 datasets by combining mRNA, miRNA, and DNA methylation (Methy) expression profiles along with their interaction networks. The synthetic datasets were created using our \texttt{omics\_GAN} model, which was designed to capture complex biological interactions across these omics layers. We evaluated the quality of the synthetic datasets for predicting Alzheimer’s Disease (AD) and cancer outcomes and for identifying significant biomarkers for these diseases. The tasks were applied to the AD patients’ ROSMAP dataset and the TCGA datasets for various cancer types. The evaluation of the synthetic datasets generated by \texttt{omics\_GAN} was conducted under the following assumptions:  
\begin{enumerate}
    \item The synthetic datasets effectively learn and preserve the expressions among mRNA, miRNA, and methylation profiles, as well as the biological interactions between them. As a result, they are expected to offer superior predictive biomarkers compared to using individual omics profiles (Omics1, Omics2, and Omics3) alone.  
    \item The availability of better predictive biomarkers will lead to more accurate disease phenotype predictions.  
\end{enumerate}

We ran the SVM classifier with the above-mentioned five-fold splitting repeated 50 times to select the best synthetic data among the five updates based on validation samples, and subsequently classified the test samples using the selected synthetic data. The average AUC scores for both the original and synthetic datasets across different conditions (ROSMAP for AD, and colon and liver cancer from TCGA datasets) are presented in Table~\ref{table:table4}. The results demonstrated that the synthetic datasets generated by \texttt{omics\_GAN} consistently outperformed the original omics profiles in almost all cases. For the mRNA profile, in the ROSMAP dataset, the AUC improved from $0.72 \pm 0.05$ (original) to $0.74 \pm 0.03$ (synthetic). Similarly, in the colon and liver cancer datasets, the AUC increased from $0.66 \pm 0.06$ to $0.77 \pm 0.04$ and from $0.68 \pm 0.04$ to $0.72 \pm 0.05$, respectively. For the miRNA profile, the synthetic datasets also showed notable improvements. In the colon dataset, the AUC rose from $0.59 \pm 0.05$ (original) to $0.69 \pm 0.03$ (synthetic), indicating enhanced predictive power.  

In the case of the Methy profile, the synthetic data exhibited superior performance. For example, in the liver dataset, the AUC increased from $0.64 \pm 0.06$ (original) to $0.71 \pm 0.05$ (synthetic).  

The results suggested that the integration of multi-omics data using \texttt{omics\_GAN} not only preserves the biological relationships between the omics profiles but also enhances the identification of predictive biomarkers. The improved AUC scores across different datasets validated the ability of \texttt{omics\_GAN} to generate high-quality synthetic data that can significantly aid in biomarker discovery and disease outcome prediction.

\begin{table}[ht]
    \centering
    \caption{Comparative analysis of the prediction performance between the original and synthetic omics profiles.}
    \label{table:table4}
    \begin{tabular}{@{}ccccc@{}}
        \toprule
    \multirow{3}{*}{{Input Values}}    & \multicolumn{2}{c}{AUC value} & \multicolumn{2}{c}{} \\ 
        \cmidrule(lr){2-5}
                     & \multicolumn{2}{c}{ROSMAP} & COLON & LIVER \\ 
        \cmidrule(lr){2-3}
                     & 245 & 351 & & \\ 
        \midrule
        mRNA           & 0.72$\pm$0.05 & 0.77$\pm$0.06 & 0.66$\pm$0.06 & 0.68$\pm$0.04 \\ 
        Synthetic-mRNA & 0.74$\pm$0.03 & 0.78$\pm$0.04 & 0.77$\pm$0.04 & 0.72$\pm$0.05 \\ 
        miRNA          & 0.68$\pm$0.07 & 0.73$\pm$0.06 & 0.59$\pm$0.05 & 0.59$\pm$0.08 \\ 
        Synthetic-miRNA & 0.73$\pm$0.06 & 0.76$\pm$0.04 & 0.69$\pm$0.03 & 0.62$\pm$0.06 \\ 
        Methy          & 0.79$\pm$0.09 & 0.77$\pm$0.08 & 0.69$\pm$0.06 & 0.64$\pm$0.06 \\ 
        Synthetic-Methy & 0.77$\pm$0.06 & 0.77$\pm$0.03 & 0.78$\pm$0.07 & 0.71$\pm$0.05 \\ 
        \bottomrule
    \end{tabular}
\end{table}

\subsection{Role of interaction network on predicting AD outcome}
The integration of Omics1, Omics2, and Omics3 into a new feature set is hypothesized to provide more comprehensive information compared to individual mRNA, miRNA, and methylation (methy) expression data. Table~4 has previously demonstrated the capability of \textit{omics\_GAN} to enhance the prediction performance for AD and cancer outcomes. We hypothesize that \textit{omics\_GAN} leverages the biological interactions among the three omics layers within a multi-omics interaction network to generate synthetic datasets with improved predictive signals. To investigate whether the performance improvement is due to the additional omics data or the model's ability to exploit the interaction network for data integration, we designed an experiment to explore the effects of the interaction network on synthetic omics data and their predictive performance.  

In this experiment, we executed the framework for Omics1, Omics2, and Omics3 using true biological networks for ROSMAP and TCGA datasets. Subsequently, we replaced the true networks with random networks that have the same density as the true ones. The prediction results for synthetic Omics1, Omics2, and Omics3 using both true and random networks are presented as boxplots in Figures~6. The results include predictions based on original and synthetics omics expression using the true and random network for ROSMAP and TCGA (Liver and Colon cancer), as illustrated in Figure~6 (A, B, C), respectively. Our observations showed that the predictive performance for synthetic data improved across Omics1, Omics2, and Omics3 in both the ROSMAP and TCGA (Liver and Colon cancer) datasets. In contrast, using a random network resulted in decreased prediction accuracy for synthetic data across the same omics layers and datasets. This suggests that the true biological interaction network plays a crucial role in enhancing the predictive performance of the synthetic omics data, highlighting the importance of accurately modeling biological interactions in multi-omics data integration.

\begin{figure*}  
    \centering
    \includegraphics[width=\textwidth]{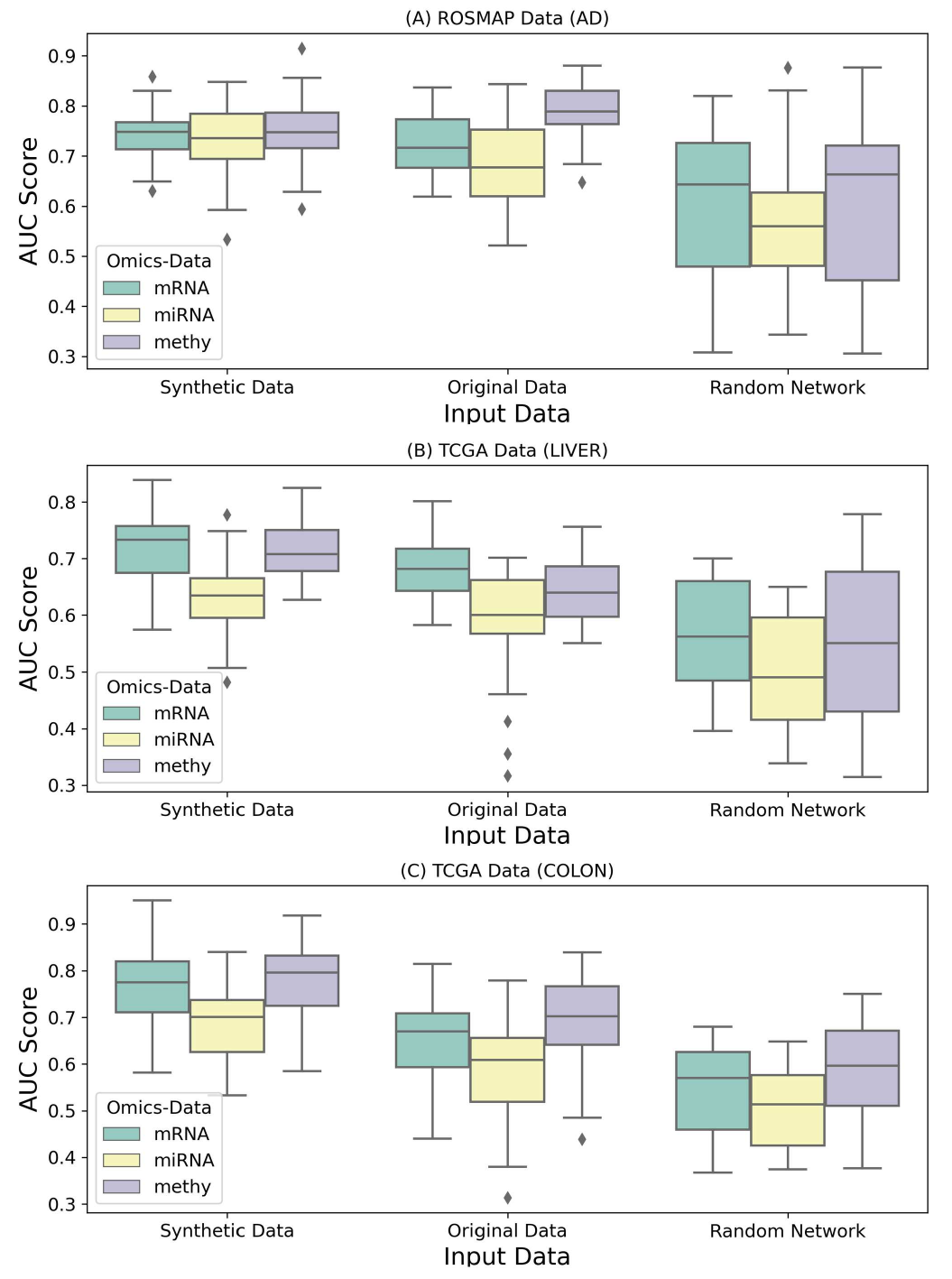} 
    \caption{Prediction results for AD and cancer patients using original, synthetic, and random networks for omics expression in ROSMAP and TCGA (Liver and Colon cancer) datasets. A indicates the prediction results for ROSMAP data. B indicates the prediction results for Liver cancer data, and C indicates the prediction results for Colon cancer data.} 
    \label{fig:Figure3}  
\end{figure*}

\subsection{Comparison Between Real and Synthetic Data}
Figures~\ref{fig:Figure3}(A, B, C),contain boxplots comparing two groups of omics data for various features across three data types: mRNA, miRNA, and methy. Each row represents a different data type, with each boxplot pair comparing real omics data (green) and generative synthetic omics data (orange). Across features (LOC284890, FU154140, and ENO3), the distributions for real data (mRNA) and synthetic data (mRNA\_G) are visually similar, with overlapping medians and ranges. mRNA\_G data slightly smoothens the spread of values compared to real data, reducing extreme outliers. For features (hsa-miR-1277, hsa-miR-424, and hsa-let-7g), synthetic data (miRNA\_G) appears to slightly shift the median while maintaining overall variability. miRNA\_G data demonstrates better uniformity in interquartile ranges (IQRs), highlighting its capacity to preserve critical miRNA signal patterns while minimizing noise. Features (cg37870146, cg19414589, and cg10648670) show more consistent distributions between real (methylation) and synthetic (methy\_G) data. methy\_G data has a reduced spread in outlier points, emphasizing its robustness. The boxplot visualizations highlighted that while real data captures authentic biological variability, synthetic data reduces noise and outliers, offering a cleaner representation of core patterns. Because, synthetic data, generated using GAN models, maintains the statistical properties of real data while ensuring robustness against experimental artifacts. By providing balanced and scalable datasets, synthetic data enhances the identification of biologically relevant features, particularly for underrepresented conditions. These advantages make synthetic data a valuable tool for advancing machine learning applications in multi-omics research and drug discovery.

\subsection{Significant Feature}
\begin{figure*}
    \centering
    \includegraphics[width=\textwidth]{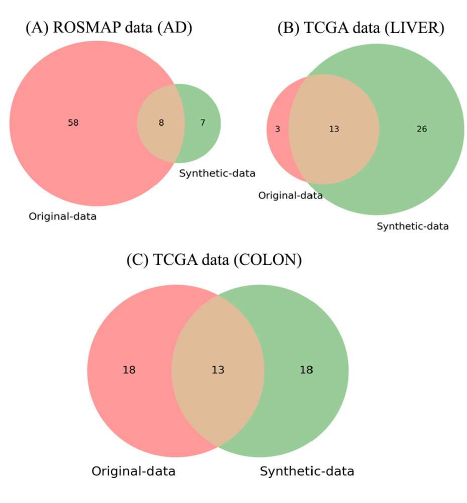} 
    \caption{Significant features of the original and synthetic datasets are visualized using a Venn diagram, where (A) represents the ROSMAP dataset for AD patients, (B) represents the TCGA liver cancer dataset, and (C) represents the TCGA colon cancer dataset.}
    \label{fig:figure7}
\end{figure*}

We conducted an in-depth analysis of the Omics\_GAN modules to identify key genes that influence disease outcomes across three omics datasets (mRNA, miRNA, and methy) from the ROSMAP and TCGA cohorts. To determine the most significant features, we applied the Wilcoxon rank-sum test with a predefined significance threshold (p-value: 0.005) to both the original and synthetic datasets. For the ROSMAP dataset, which is associated with AD, our analysis revealed 66 significant genes in the original dataset and 15 significant genes in the synthetic dataset. From these, we identified 8 common genes (METTL1, ANLN, ADIPOR2, NRIP2, RIPOR3, SAMD4A, PLEKHB1, and CLCA4) that are particularly relevant for drug repurposing in AD patients (Figure~\ref{fig:figure7}A). For liver cancer (TCGA dataset), we identified 16 significant genes in the original dataset and 39 significant genes in the synthetic dataset. We selected 13 common genes (SLC7A8, C1orf212, MAP7D3, S100A9, HPS3, FFAR3, SOX15, TFF2, ZG16B, PTP4A2, C17orf44, SH3GLB1, EIF5B) for drug repurposing in liver cancer (Figure~\ref{fig:figure7}B). For colon cancer (TCGA dataset), our analysis revealed 31 significant genes in both the original and synthetic datasets. We selected 13 common genes (AMIGO1, ANKAR, CCDC47, EGFL8, ENO3, FLJ45244, LEKR1, LOC100272217, LOC202781, LOC284900, MTHFSD, NGRN, RPS24) for drug repurposing in colon cancer (Figure~\ref{fig:figure7}C). A complete list of all significant genes identified in our analysis is provided in Supplementary Table 1-6.

\subsection{GO Functions and KEGG Pathway Enrichment Analysis}

Gene Ontology (GO) functional analysis and Kyoto Encyclopedia of Genes and Genomes (KEGG) pathway enrichment analysis~\cite{80,81} are widely applied approaches for identifying significantly enriched biological functions and pathways associated with differentially expressed or significant genes. These analyses are essential for uncovering the molecular mechanisms and cellular roles of genes. 

In this study, GO and KEGG enrichment analyses were performed using the Enrichr web tool (\url{https://maayanlab.cloud/Enrichr/}, accessed on 12 October 2025). A significance threshold of $p < 0.05$ was applied. The top five significantly enriched GO terms and KEGG pathways for each disease are summarized in Table~\ref{table:table5}. For Alzheimer’s disease (AD), the top enriched GO terms included Vesicle Transport Along Microtubule, Vesicle Cytoskeletal Trafficking, Sodium Ion Transport, Organelle Transport Along Microtubule, and Lysosome Localization. The corresponding KEGG pathways were PtdIns 4,5 P2 in Cytokinesis (WP5199); Development of Pulmonary Dendritic Cells and Macrophage Subsets (WP3892); Biosynthesis of Tetrahydrobiopterin and Phenylalanine Catabolism (WP4156); MAPK Pathway in Congenital Thyroid Cancer (WP4928); and Hedgehog Signaling (WP47). For liver cancer, the top enriched GO terms included Regulation of Chemokine Production, Response to Amino Acid Starvation, Cellular Response to Amino Acid Starvation, Positive Regulation of Chemokine Production, and Regulation of Respiratory Burst Involved in Inflammatory Response. The corresponding KEGG pathways were Translation Factors (WP107); IL10 Anti-Inflammatory Signaling (WP4495); NRF2 ARE Regulation (WP4357); Photodynamic Therapy Induced NFE2L2/NRF2 Survival Signaling (WP3612); and Unfolded Protein Response (WP4925). For colon cancer, the top enriched GO terms included Iron Ion Transport, Regulation of Small GTPase-Mediated Signal Transduction, Rho Protein Signal Transduction, Negative Regulation of Lipid Transport, and Regulation of Hyaluronan Biosynthetic Process. The significantly enriched KEGG pathways included Integrin-Mediated Cell Adhesion (WP185); Osteopontin Signaling (WP1434); and Mevalonate Arm of Cholesterol Biosynthesis (WP4190). Additional significantly enriched GO terms and KEGG pathways are provided in Supplementary Tables 7–9.

\begin{table*}[htbp]

\caption{Summary of the top five significantly enriched (p < 0.05) Gene Ontology (GO) terms and KEGG pathways derived from significant genes in the synthetic and original datasets.}
\resizebox{\textwidth}{!}{%
\begin{tabular}{>{\centering\arraybackslash}p{5.5cm}
                >{\centering\arraybackslash}p{2.2cm}
                >{\centering\arraybackslash}p{2.8cm}
                >{\centering\arraybackslash}p{5.5cm}}
\hline
\textbf{Term} & \textbf{P-value} & \textbf{Combined d Score} & \textbf{Genes} \\
\hline
\multicolumn{4}{c}{\textbf{Biological Process for AD}} \\
\hline
Vesicle Transport Along Microtubule (GO:0047496) & 0.003171 & 153.1356 & NDE1; KIF5B \\
Vesicle Cytoskeletal Trafficking (GO:0099518) & 0.003439 & 144.405 & KIF5B; NDE1 \\
Sodium Ion Transport (GO:0006814) & 0.004649 & 51.2976 & SLC6A9; SLC6A12; SLC4A11 \\
Organelle Transport Along Microtubule (GO:0072384) & 0.009067 & 70.55036 & KIF5B; NDE1 \\
Lysosome Localization (GO:0032418) & 0.009499 & 68.10245 & KIF5B; PLEKHM2 \\
\hline
\multicolumn{4}{c}{\textbf{Biological Process for Liver}} \\
\hline
Regulation of Chemokine Production (GO:0032642) & 0.003681 & 136.1289 & FFAR3; HMOX1 \\
Response to Amino Acid Starvation (GO:1990928) & 0.003851 & 131.8095 & SH3GLB1; EIF2S1 \\
Cellular Response to Amino Acid Starvation (GO:0034198) & 0.004383 & 120.1402 & SH3GLB1; EIF2S1 \\
Positive Regulation of Chemokine Production (GO:0032722) & 0.005749 & 98.73995 & FFAR3; HMOX1 \\
Regulation of Respiratory Burst Involved in Inflammatory Response (GO:0060264) & 0.010457 & 554.8782 & S100A9 \\
\hline
\multicolumn{4}{c}{\textbf{Biological Process for Colon}} \\
\hline
Iron Ion Transport (GO:0006826) & 0.001086 & 321.6182 & SLC41A2; CLTC \\
Regulation of Small GTPase Mediated Signal Transduction (GO:0051056) & 0.006713 & 41.95023 & GNA13; ARHGAP4; VAV2 \\
Rho Protein Signal Transduction (GO:0007266) & 0.008039 & 77.05991 & GNA13; ARHGAP4 \\
Negative Regulation of Lipid Transport (GO:0032369) & 0.012191 & 457.8501 & ITGB3 \\
Regulation of Hyaluronan Biosynthetic Process (GO:1900125) & 0.012191 & 457.8501 & CLTC \\
\hline
\multicolumn{4}{c}{\textbf{KEGG Pathway for AD}} \\
\hline
PtdIns 4 5 P2 in Cytokinesis Pathway WP5199 & 0.041212 & 83.69212 & ANLN \\
Development of Pulmonary Dendritic Cells And Macrophage Subsets WP3892 & 0.04457 & 74.82974 & FLT3LG \\
\hline
\end{tabular}%
}
\label{table:table5}
\end{table*}

\begin{figure*}
    \centering
    \includegraphics[width=\textwidth]{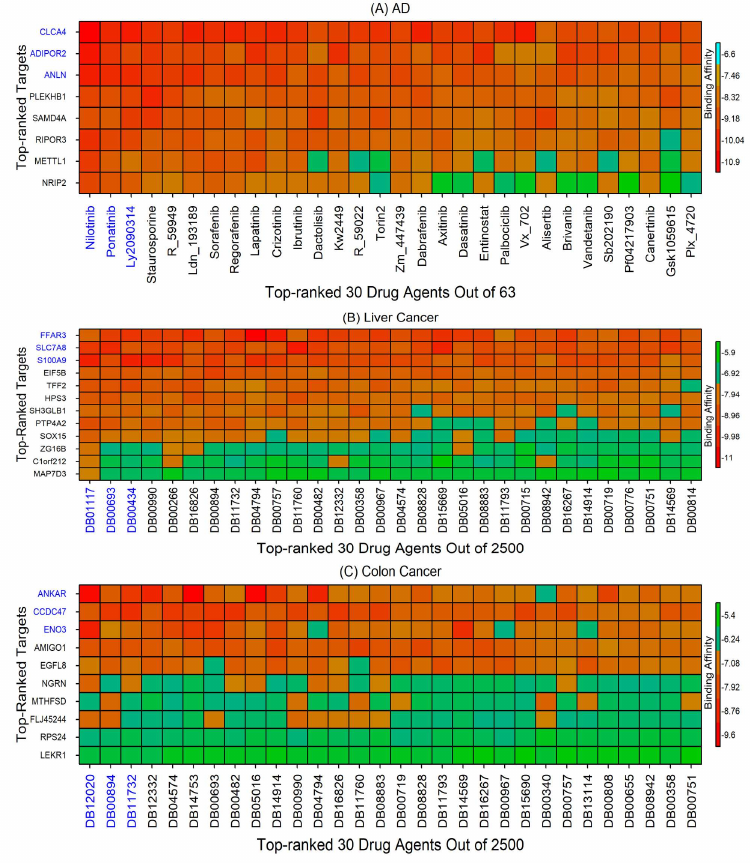} 
    \caption{Image of binding affinities based on the top-ordered 30 drug agents against the ordered target proteins, where red colors indicated the strong binding affinities. (A) represents the binding score for AD patients, (B) represents the binding score for liver cancer, and (C) represents the binding score for colon cancer.
}
    \label{fig:figure8}
\end{figure*}
\begin{figure*}
    \centering
    \includegraphics[width=\textwidth]{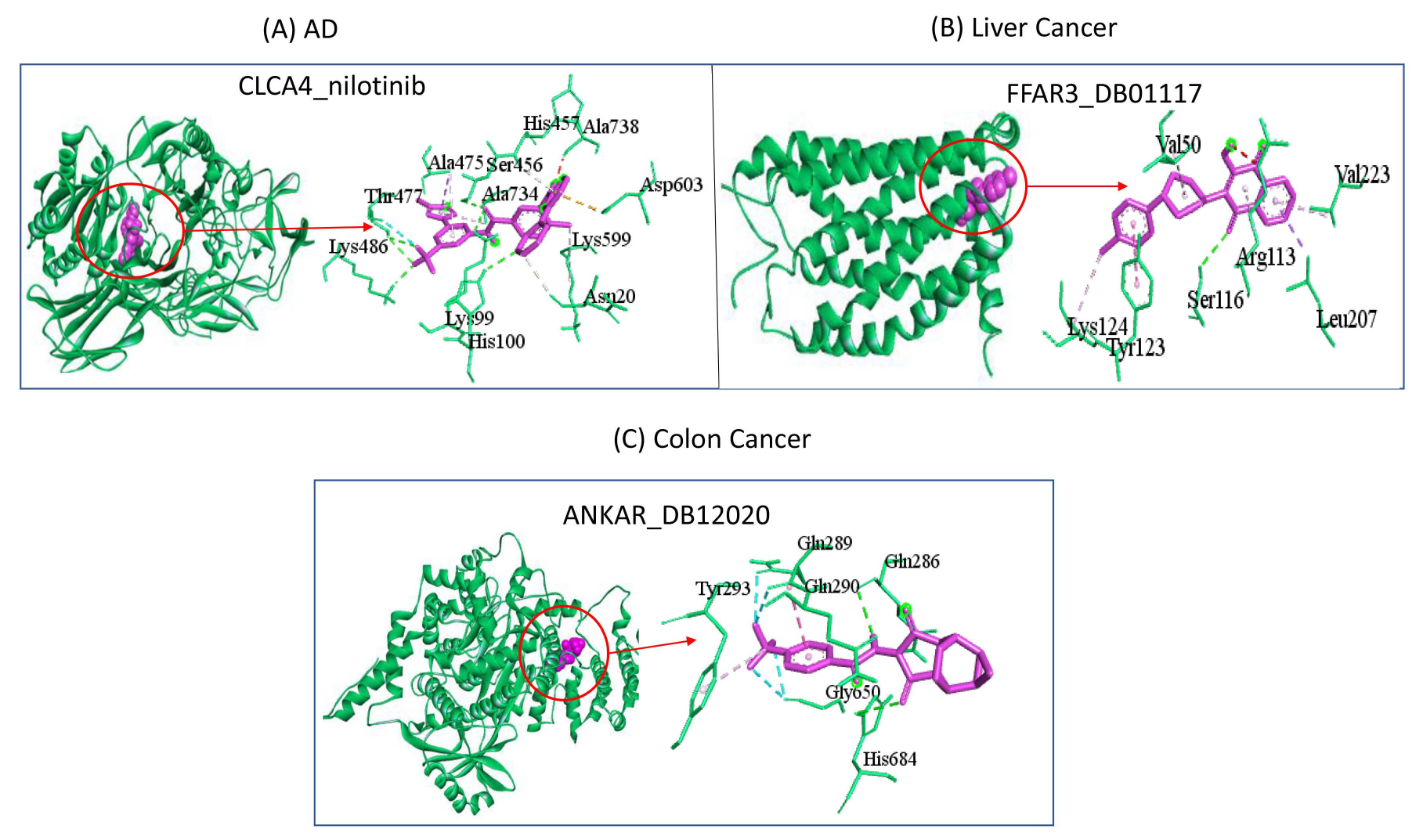} 
    \caption{The top-ranked three complexes obtained from molecular docking study for AD, Liver and colon cancer and their two-dimensional chemical interactions. The figure is generated by using discovery studio visualizers software. Complexes: (A) represents the CLCA4–Nilotinib complex for AD patients, (B) represents the FFAR3–DB01117 complex for liver cancer, and (C) represents the ANKAR–DB12020 complex for colon cancer.
    }
    \label{fig:figure9}
\end{figure*}

\subsection{Identification of Effective Drug Candidates for Disease Outcomes}
In this study, we explored potential therapeutic drug candidates for Alzheimer's disease (AD), liver cancer, and colon cancer using molecular docking simulations. The objective was to systematically assess drug-protein interactions and identify promising compounds based on their binding affinity scores.

\textbf{Alzheimer’s Disease} \\
To identify potential drug candidates for AD through molecular docking simulations, we considered eight drug target proteins and 63 meta-drug agents, as detailed in the data sources. The 3D structures of the METTL1, ANLN, and ADIPOR2 proteins were obtained from the Protein Data Bank (PDB)~\cite{ProtineBank} with source codes 3ckk, 2y7b, and 5lwy, respectively. Additionally, the NRIP2, RIPOR3, SAMD4A, PLEKHB1, and CLCA4 proteins were retrieved from AlphaFold via the UniProt Webserver~\cite{UniPort}. The 3D structures of 63 drug agents were downloaded from the PubChem database~\cite{PubChem}. 

Molecular docking was performed between 8 proteins and 63 drug agents to compute the binding affinity scores (kcal/mol) for each protein–drug pair. The proteins were ranked in descending order based on the row sums of the binding affinity matrix, while drug agents were ranked according to the column sums. Subsequently, the top three drug agents—Nilotinib, Ponatinib, and Ly2090314—were identified as potential candidates, all exhibiting average binding affinity scores of $\leq -8.00$ kcal/mol against the top-ranked 35 drugs (see Figure~8A). The top-ranked virtual hit complex (CLCA4–Nilotinib) from AutoDock-Vina docking was further analyzed for protein–ligand interactions. As illustrated in Figure~9A, the CLCA4–Nilotinib complex exhibited six hydrogen bonds with Asn20, Lys99, His100, Ser456, Thr477, and Lys486 residues. Additionally, significant hydrophobic interactions were observed with Ala474, Ala475, and Lys599 residues. Further electrostatic and halogen interactions were noted with Asp603 and Thr477, respectively. Nilotinib (AMN107) is a tyrosine kinase inhibitor under investigation for its potential in treating chronic myelogenous leukemia (CML) (DrugBank). Based on our findings, Nilotinib may play a crucial role in the treatment of AD. Additional docking scores are provided in Supplementary Table~10.

\textbf{Liver Cancer} \\
For liver cancer, we conducted molecular docking simulations to identify drug candidates, considering 13 target proteins and 2,500 DrugBank agents, as specified in the data sources. The 3D structures of FFAR3, SLC7A8, S100A9, EIF5B, SH3GLB1, PTP4A2, and ZG16B were retrieved from the Protein Data Bank (PDB) with source codes 8j20, 7cmi, 4ggf, 7tql, 9g2u, 5k22, and 3aqg, respectively. The TFF2, HPS3, SOX15, C1orf212, and MAP7D3 proteins were obtained from AlphaFold, while C17orf44 did not have an available 3D protein structure. The 3D structures of 2,500 FDA-approved drugs were downloaded from DrugBank.

Molecular docking was conducted between 12 proteins and 2,500 drug agents, generating a binding affinity matrix. The proteins and drug agents were ranked similarly to the AD analysis. The top three drug agents—DB01117, DB00693, and DB00434—were selected based on their average binding affinity scores of $\leq -8.00$ kcal/mol (see Figure~8B). The FFAR3–DB01117 complex, a top-ranked interaction, was further analyzed for protein–ligand binding characteristics (Figure~9B). The complex exhibited one hydrogen bond with the Ser116 residue, along with major hydrophobic interactions involving Leu207, Tyr123, Val50, Lys124, Arg113, and Val223 residues. DB01117, known as Atovaquone, is a hydroxynaphthoquinone with antimicrobial and antipneumocystis activity, primarily used in antimalarial protocols (DrugBank). These findings suggest that Atovaquone (DB01117) may be a promising candidate for liver cancer treatment. Supplementary Table~11 contains additional docking scores.

\textbf{Colon Cancer} \\
For colon cancer, we applied molecular docking simulations to identify candidate drugs, considering 13 target proteins and 2,500 DrugBank agents, as outlined in the data sources. The 3D structures of ENO3 and MTHFSD were retrieved from the Protein Data Bank (PDB) with source codes 2xsx and 2e5j, respectively. The ANKAR, CCDC47, AMIGO1, EGFL8, NGRN, FLJ45244, RPS24, and LEKR1 proteins were sourced from AlphaFold, while the LOC100272217, LOC202781, and LOC284900 protein structures were unavailable.

Molecular docking was performed between 10 proteins and 2,500 drug agents, and ranking was conducted similarly to previous cases. The top three drug agents (DB12020, DB00894, and DB11732) were identified based on their average binding affinity scores of $\leq -8.00$ kcal/mol (see Figure~8C). Further protein–ligand interaction profiling of the ANKAR–DB12020 complex (Figure~9C) revealed two hydrogen bonds with His684 and Gln286 residues. Additionally, hydrophobic interactions were observed with Gln289, Gln290, and Tyr293 residues, while electrostatic and halogen interactions were noted with Asp603 and (Gln289, Gly650), respectively. DB12020, known as Tecovirimat, was approved by the FDA in July 2018 as the first drug indicated for smallpox treatment (\url{https://www.fda.gov/news-events/press-announcements/fda-approves-first-drug-indication-treatment-smallpox}). It is available in both oral and intravenous formulations (DrugBank). Based on these findings, Tecovirimat may have therapeutic potential for colon cancer treatment. Further docking score results are presented in Supplementary Table~12.

\section{Conclusion}
\label{Sec:dis}
This study demonstrates that Omics\_GAN is an effective tool for generating high-quality synthetic multi-omics data, which can enhance disease outcome predictions and aid in biomarker discovery. The results highlight the significant role of biological interaction networks in optimizing the predictive performance of synthetic data, with implications for drug repurposing. Additionally, molecular docking simulations were used to identify key therapeutic drug candidates for Alzheimer's disease, liver cancer, and colon cancer. By utilizing mRNA, miRNA, and methylation datasets from the ROSMAP and TCGA cohorts, we identified significant genes that are relevant for drug repurposing in these diseases, providing valuable insights for future therapeutic strategies. The docking simulations further identified promising drug candidates, such as Nilotinib for Alzheimer's disease, Atovaquone for liver cancer, and Tecovirimat for colon cancer, all of which demonstrated strong binding affinities and specific interactions with key target proteins. These findings emphasize the importance of multi-omics integration and computational drug discovery in advancing precision medicine and uncovering novel therapeutic options for complex diseases.

\bibliographystyle{plainnat}
\bibliography{Ref-CI-2021}
\end{document}